# Wettability and Color Change of Copper by Controlling Area Fraction of Laser Ablated Surface

Mantas Gaidys[1], Stella Maragkaki[2], Alexandros Mimidis[2], Antonis Papadopoulos[2], Andreas Lemonis[2], Evangelos Skoulas[2], Andrius Žemaitis[1], Emmanuel Stratakis[2], Mindaugas Gedvilas[1]*

[1]Department of Laser Technologies (LTS), Center for Physical Sciences and Technology (FTMC), Savanoriu Ave. 231, 02300 Vilnius, Lithuania

[2]Institute of Electronic Structure and Laser (IESL), Foundation for Research and Technology (FORTH), N. Plastira 100, Vassilika Vouton, 70013 Heraklion, Crete, Greece

E-mail: mindaugas.gedvilas@ftmc.lt

**Abstract**

In this research, wettability control by area fraction of laser ablated surface of copper is presented. The functional surfaces with full wettability control from highly hydrophilic to super-hydrophobic were created on copper by nanosecond (ns) and picosecond (ps) laser irradiation. The area fraction and color change were evaluated by digital image processing of microscopic images of the laser-ablated copper surface. The control of the wetting angle from almost 0 degrees to 132 degrees was achieved for both ps and ns pulses by controlling the area fraction of the laser-ablated surface. Cassie, Cassie-Baxter, and Wenzel models were adopted to explain the experimental results. For the first time, the wettability and color of copper were controlled by controlling the area fraction of the laser-ablated surface. It is expected that the current results make an impact on the heat exchanger technology of water heat sinks, cooling units, atmospheric water generators, and fog harvesting and impact numerous applications from power plants to solar thermal water systems devices where highly-hydrophilic to super-hydrophobic copper can be applied.

**Keywords**

Super-hydrophobic, highly-hydrophilic, area fraction, copper, laser ablation, surface roughness, color change.

## 1. Introduction

A wide variety of biological breeds found in living nature that have superb functional surface properties include birds [1], plants [2-5], insects [6-10], fish [11], sea mammals [12,13], and reptiles [14]. These unique functionalities like hydrophilicity [15], fog-harvesting [16], structural coloration [17], antibacterial [18], super-hydrophobicity [19], self-cleaning [20], anti-fogging [21], anti-reflective [22], drag-reducing [23], etc. have evolved over millions of years to extremes and provide those species with the ability to live and survive, and further develop new multi-functions to adapt to the fast-changing environment. The genetic modification of breeds due to natural selection has helped to evolve the specific surface functionalities. In most cases, the unique surface functions come from their morphology in the mesoscale, with the tri-modal dimensions from nano to micro and macro [12,24,25].

The scientific branches such as bionics, nature/bio-inspiration, bio-replication, bio-engineering, and bio-mimetics have evolved during the last two decades to copy excellent surfaces from nature [26-35]. Direct laser writing is a precision micro-fabrication technology with large flexibility options [36-38]. The fabrication of tri-modal structures with the nano-, micro-, and macro-length scales in one surface can be achieved by this technique [39,40]. Moreover, fabrication using ultra-short laser pulses has demonstrated unique features such as efficient ablation [11,41], cold-ablation [42,43], ablation-cooled efficient material removal [44], and bi-stable ablation [45] which enables high precision repeatable texturing for a large versatility of materials. Laser beam interference ablation is a flexible tool for the periodical structuring [46,47]. However, minimal feature size in the sub-micro-scale is restricted by the diffraction limit, which is in the order of irradiation wavelength. On the other hand, the bio-replication of functional surfaces usually requires laser processing beyond the diffraction limit at the nanoscale [27,29]. The diffraction limit can be beaten by reaching a nanoscale resolution with laser-induced periodic surface structures (LIPSS)[48], which can be textured on a large scale with a high rate [49-51].

The super-hydrophobic copper structuring by laser has been reported in numerous scientific works using nanosecond [52-55], picosecond [56,57], and femtosecond [58,59] lasers. However, making the opposite, super-hydrophilic copper, using laser texturing is still an open question. There is no scientific work found in literature, where perfect wetting of copper surface would be achieved by laser irradiation. However, there are other competing techniques capable of producing super-hydrophilic copper, like, electrochemical deposition [60,61], layer-by-layer self-assembly [62], one-step liquidus modification [63], etc.

Highly wetting textured surfaces have shown promise in boiling applications since capillary increases the maximum heat flux that can be dissipated [64-66], which has a huge potential in the heat exchanger technology of water heat sinks [67], fans, and cooling units [67,68]. Also, super-wetting surfaces can be applied as fog-harvesting systems [69-72] that can dramatically improve atmospheric water generator performances [73,74]. The current cutting-edge preparation techniques of super-hydrophilic and super-hydrophobic surfaces require time-consuming processes [61] and complex multiple-steps [60-62], or processes that produce chemically hazardous wastes [75,76]. At the same time, the mechanical durability [77], gradual degradation due to long exposure to outdoor conditions [78], and degradation in time [79] of chemically replicated bio-inspired surfaces in many cases are unsolved problems [80,81]. The laser functionalization of the surface by texturing using ultrashort laser pulses is simple, low-cost, and chemical-free. It can be easily scaled up using commercially available industrial laser-processing systems [82,83]. The achieved functional properties and surface morphologies of the laser-fabricated textures were found close to the leaf of the cactus [69,84] and the underwater side of the lily leaf [85,86]. The biomimetic fabrication using laser irradiation considering the simplicity of the process and high processing rate together with the robustness of achieved super-wetting surfaces can be applied in power plants [74] and solar thermal water systems [87].

In this work, we demonstrate a novel, single-step, chemical-free fabrication method for producing super-wetting and highly hydrophobic copper surfaces using ns and ps lasers. For the first time, simultaneous control of the wettability and color of copper is achieved by systematically varying the area fraction of the laser-ablated surface, offering a versatile approach for surface modification. Our study adopted the Cassie, Cassie-Baxter, and Wenzel models to explain that as the ablated area fraction increases, the contact angle decreases, enhancing hydrophilicity. Furthermore, significant visual and optical changes such as variations in color distance, gray value, and gray luminance correlate linearly with the ablated area fraction, showcasing the transformative impact of laser parameters. Shorter pulse durations (10 ps) demonstrate a particularly stronger effect than longer (10 ns), underscoring the potential of this technique for advanced material applications.

## 2.   Theoretical background

The characterization of the wetting properties of a surface is defined on the static contact angle measured by a sessile droplet technique [88]. For water droplets, a surface having a contact angle smaller than 90° is hydrophilic, while one larger than 90° is hydrophobic. The Wenzel, Cassie, and Cassie-Baxter models are widely used in the research of wetting behavior and surface interactions, particularly in the context of hydrophobicity and super-hydrophobicity.

On textured rough surfaces with super-hydrophilic properties, the liquid spreads completely and a near-zero contact angle is achieved. Such super-hydrophilic transition of roughened surface is explained by the Wenzel model [89], where the static contact angle in this Wenzel state is smaller than one on a flat hydrophilic surface of the same material.

Contrarily, textured hydrophobic surfaces can provide very different scenarios, depending on which of two distinct wetting states is attained. Super-hydrophobic case, defined by a static contact angle exceeding 150° having a roll-off angle less than 10°, is explained by the Cassie-Baxter model [90] in which air remains trapped inside the texture, causing a liquid to sit on both air and solid. An alternative, super-hydrophobic case is explained in the Wenzel model [89], where the liquid fills the surface structures without air trapped under the water. The static contact angle in this Wenzel state is larger than one on a flat hydrophobic surface of the same material, but typically does not exceed 150°, as it lacks the air trapping characteristic of the Cassie-Baxter state required for superhydrophobicity.

These models are fundamental to understanding how surface textures can influence wetting behavior, therefore the next sub-sections are dedicated to a brief introduction to those classical models.

## 2.1. Wenzel Model:

This Wenzel model assumes that the liquid completely penetrates the rough surface. The apparent contact angle is a function of the intrinsic contact angle and the surface roughness: [89]

$$\cos\theta^* = r_1 \cos\theta_1, \tag{1}$$

where $\theta^*$ is the measured contact angle on the rough surface, $r_1$ is the ratio of actual surface area to projected surface area, and $\theta_1$ is the intrinsic contact angle on a smooth surface. Wenzel state has a larger surface wettability because the liquid on the surface enters the grooves, increasing the contact area.

## 2.2. Cassie Model:

This model addresses surfaces with regions of different wettability coexist caused by surface chemical heterogeneity: [91,92]

$$\cos\theta^* = f_1 \cos\theta_1 + f_2 \cos\theta_2, \tag{2}$$

where $f_1$ and $f_2$ are the fractions of the surface with different wettability angles $\theta_1$ and $\theta_2$, respectively. The fractional areas of two different surface components fulfill the requirement for the sum of the fractions must be equal as $f_1 + f_2 = 1$, this constraint ensures that no overlapping or missing regions exist.

## 2.3. Cassie-Baxter Model:

When the composite contact surface is composed of air and solid since the contact angle of liquid and air is 180°, Eq. (2) can be simplified to [93,94]:

$$\cos\theta^* = f_1(1+\cos\theta_1) - 1. \tag{3}$$

where $\theta^*$ is the measured contact angle on a composite surface, $f_1$ is the fraction of the solid-liquid interface area, while the rest fraction $(1 - f_1)$ is air, $\theta_1$ is the intrinsic contact angle. This model assumes that the liquid rests on top of surface roughness features, with air pockets trapped underneath. This creates a composite surface of solid and air, which can drastically increase hydrophobicity.

In scientific literature, the terms Cassie-Baxter and Cassie are commonly used. Cassie-Baxter is used to refer to uneven surfaces incompletely wetted by a liquid when vapor remains under the drop. Cassie is used only for even or rough surfaces of a solid state that is completely wetted by the liquid.

## 2.4. Wenzel and Cassie-Baxter model:

The wetting angle on rough and heterogeneous surfaces is described using the combined Wenzel and Cassie-Baxter models: [94,95]

$$\cos\theta^* = r_1 f_1 \cos\theta_1 + f_2 \cos\theta_2, \tag{4}$$

which considers surface roughness, where the Wenzel part of Eq. (1) with roughness factor $r_1$ amplifies the wetting properties (hydrophilic or hydrophobic) and Cassie-Baxter's part of Eq. (2) accounts for surface heterogeneity, using a fractional areas $f_1$ and $f_2$ of different surface components to adjust the contact angle.

## 3. Experimental

### 3.1. Sample preparation

The samples used for laser texturing and contact angle measurements were square copper substrates (CW004A, Ekstremalė) with dimensions of 50 × 50 mm and a thickness of 5 mm, featuring a purity of 99.9%. These samples had a mirror finish with a surface roughness of approximately Ra ~ 6 nm, measured using a 3D optical profiler (S neox, Sensofar).

### 3.2. Laser structuring

The principal experimental scheme of laser processing using an irradiation source with optional pulse durations and a galvanometer scanner is depicted in Figure 1.

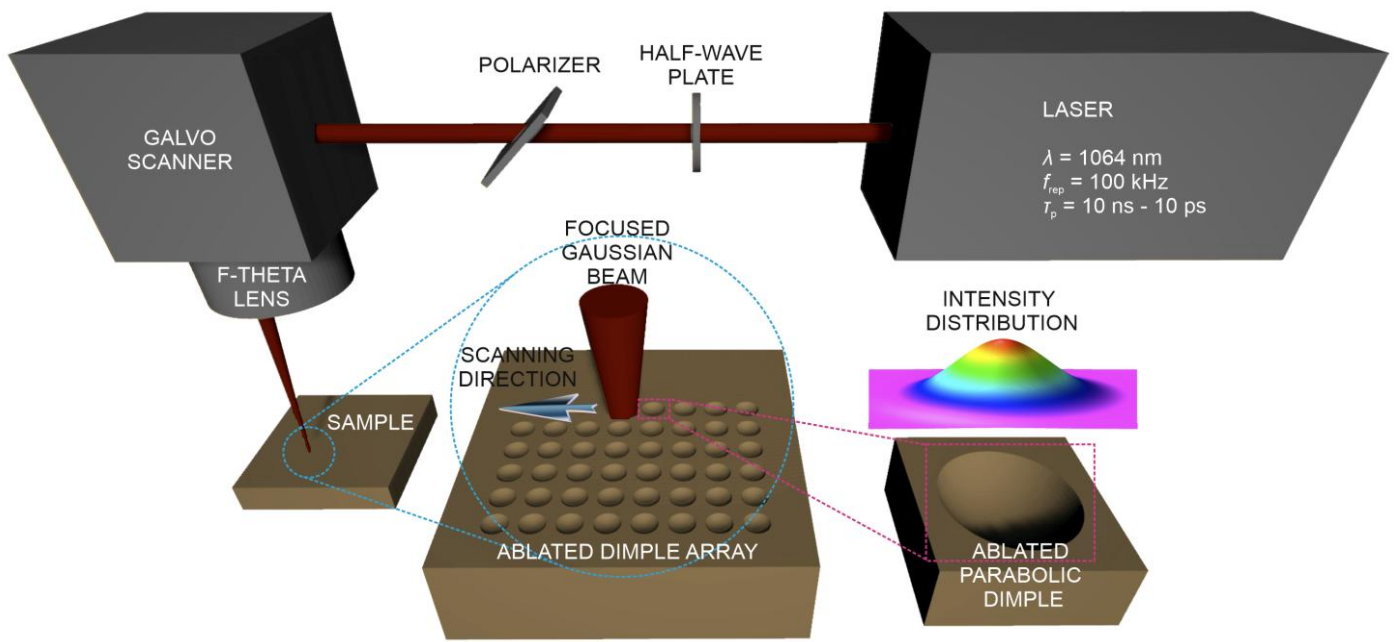

**Figure 1.** Principal scheme of laser structuring experimental setup: laser irradiation source with optional pulse durations of $\tau_p$ = 10 ns and $\tau_p$ = 10 ps, the repetition rate of $f_{rep}$ = 100 kHz, irradiation wavelength of $\lambda$ = 1064 nm, and pulse energy up to $E_p$ = 60 μJ. Beam attenuation for energy control consists of a half-waveplate and the polarizer, a galvanometric scanner equipped with a telecentric f-theta lens for Gaussian beam focus on the surface of the copper sample. The enlarged area in the middle shows a two-dimensional array of dimples ablated by scanning a Gaussian beam on the sample. The enlarged area on the bottom right corner shows a parabolic dimple ablated by a single laser pulse of a Gaussian beam intensity distribution.

Laser structuring experiments were conducted using two laser irradiation sources (Baltic HP, Ekspla, and Atlantic HE, Ekspla) with different pulse durations of $\tau_p$ = 10 ns and $\tau_p$ = 10 ps, respectively. Both lasers provided light pulses with pulse energy from $E_p$ = 1 µJ up to $E_p$ = 60 µJ at the repetition rate of $f_{rep}$ = 100 kHz, an average power of 6.0 W, and irradiation wavelengths of $\lambda$ = 1064 nm. The beam position on the metal sample surface was controlled by using a galvanometer scanner (Scangine 14, Scanlab) and scanner application software (SAMLight, SCAPS). Translation of the laser spot on the target material at a controllable speed up to $v_{scan}$ = 1.0 m/s provided the controllable distances between the transverse irradiation spots and distances between bidirectional scanned lines. The telecentric f-theta objective lens with a focal length of 80 mm was used to focus the beam on the surface of the target material. The array of rectangular areas with transverse spatial dimensions of 11.5 × 12 mm² was laser textured in 29.9 s processing time including all scanner delays. The texturing rate of ~ 5 mm²/s excluding delays of the scanner was achieved for the area of the whole scanner field of 60 × 60 mm². The path of the scanned beam on the copper sample followed a snake-like trajectory consisting of parallel lines of overlapped laser pulses. The beam was scanned along the horizontal axis at a speed of $v_{scan}$ = 1.0 m/s. The distance between consecutive laser pulses was $\Delta x$ = $v_{scan}$ / $f_{rep}$ = 10 µm. A slower translation of the beam, at a speed of 10 mm/s, was applied along the vertical axis. The distance between scanned lines in the vertical direction (hatch) was $\Delta y$ = 5 µm. The polarization of the beams was oriented along the vertical axis. The laser power was changed during the test from 0.1 W to 0.9 W (step 0.1 W, 9 tests) and from 1.0 W to 6.0 W (step 1.0 W, 6 tests) which provided controllable laser fluence on the sample. The 15 rectangular squares were marked using each laser at different laser powers which provided different laser fluence on the samples. The 15 different values of laser fluences were used, from 0.16 J/cm² to 1.44 J/cm² (step 0.16 J/cm², 9 tests) and from 1.6 J/cm² to 9.6 J/cm² (step 1.6 J/cm², 6 tests). Laser untreated copper is declared in graphs at the laser output power of 0.0 W and corresponding laser fluence of 0.0 J/cm².

Figure 2(a) illustrates the fluence-dependant morphological change of copper surface by ablation of a two-dimensional array of parabolic dimples by bidirectional scanning of a Gaussian beam.

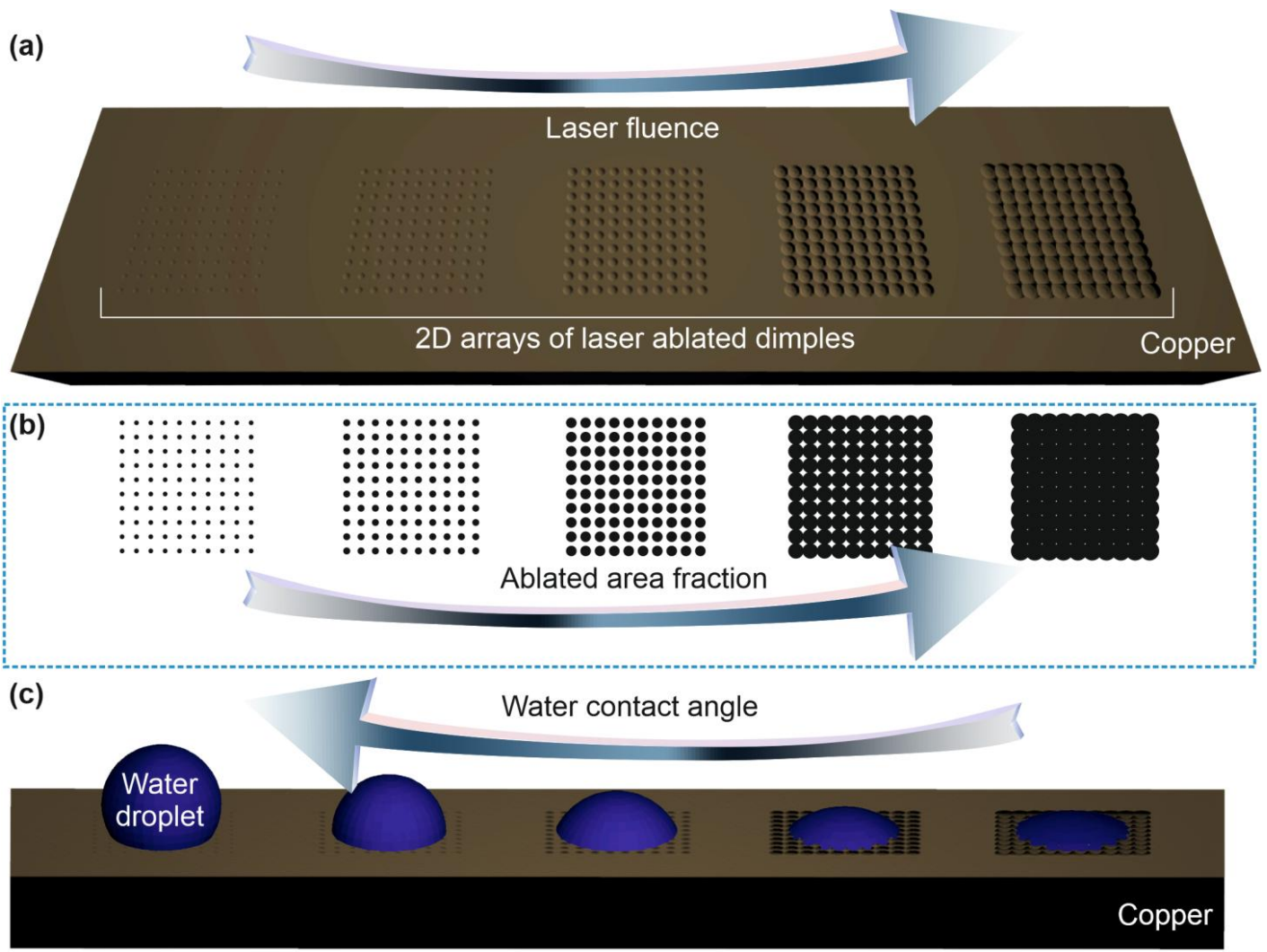

**Figure 2.** Schematic Illustration of ablated area fraction dependence on laser fluence and its influence on water contact angle: (a) five laser scanned areas consisting of a two-dimensional array of laser ablated dimples and its characteristic surface morphology depending on the laser fluence; (b) schematic representation of laser ablated areas increase with increasing laser fluence; (c) characteristic water contact angle decrease with increasing ablated area fraction and laser fluence.

The schematic representation transverse geometric laser ablated area of the copper is depicted in Figure 2(b). The ablated area fraction increases with the growth of laser fluence. Figure 2(c) illustrates ablated area fraction and fluence-dependant water contact angle. The water contact angle tends to decrease with increasing ablated area fraction and laser fluence.

### 3.3. Fluence characterization

The radii of the transverse-focused beam spots on the surface of copper were measured by the Liu ($D^2$) method described in [96]. The dimples were ablated using different pulse energies. Then, the crater size dependence on pulse energy was analyzed and beam spot radii on the sample were retrieved for both irradiation sources. For size characterization of ablated craters (dimples) on copper at different pulse energies an optical microscope (Eclipse LV100, Nikon) was used. The microscope was equipped with

5-megapixel charge-coupled device camera (DS-Fi1, Nikon), camera controller (Digital Sight DS-U2, Nikon), microscope objective (LU Plan Fluor 20x, Nikon), halogen lamp (LV-HL50PC, Nikon), and image processing software (NIS-Elements D, Nikon). The measured laser spot radii on copper were ($w_0$ = 20 ± 1 µm) equal for both nanosecond and picosecond lasers. The laser spot sizes on the sample were not changed during the tests. The power meter (Nova II, Ophir) equipped with a thermal power sensor (30A-BB-18, Ophir) was employed to measure the average laser power.

### 3.4. Contact angle measurements

The contact angle characterization was performed using an optical system designed for contact angle and contour analysis (OCA 15EC, Data-Physics Instruments GmbH) equipped with a charge-coupled device camera to capture lateral images of water droplets applied on laser textured and untreated copper. A droplet of deionized water with a volume of 3 µL was deposited by using a pipette (1 - 10 µJ Transferpette, Brand). The camera, the water droplet, and the illumination source equipped with a light-emitting diode are perfectly aligned, therefore, the droplet shadow is projected and captured by the digital camera. Average values of contact angle and measurement errors were obtained by software (SCA20, Data-Physics Instruments GmbH). Since it is known that the contact angle of distilled water and laser-irradiated metal surfaces changes approximately 10 days after the irradiation and then reaches a stable value [97,98], all the measurements presented in this work were conducted 15 days after the laser texturing.

### 3.5. Roughness evaluation

3D optical profiler (S neox, Sensofar) was employed for 3D surface topography imaging of laser ablated copper surface.

### 3.6. Profile arc length evaluation.

The normalized profile arc length $r_1$, which is the ratio of actual profile length to projected length was evaluated from height $h = h(x)$ profiles of laser textured copper by the equation:

$$r_1 = \frac{\int_{x_1}^{x_2}\sqrt{1+(dh/dx)^2}\,dx}{x_2 - x_1}. \tag{5}$$

The top part of the equation refers to the arc length of the curve, which was calculated using the arc length formula, the bottom part of the equation refers to the straight-line distance between the start $x_1$ and end points $x_2$ of the curve, projected onto horizontal axis.

### 3.7. Optical microscope photographing

Digital images of both laser-processed and untreated copper surfaces were captured using an optical microscope (Nikon Eclipse LV100) equipped with a 5-megapixel high-definition CCD camera (Nikon

DS-Fi1) with a resolution of 2560 × 1920 pixels. The camera was controlled via a Digital Sight DS-U2 controller and NIS-Elements D imaging software, both from Nikon. For imaging, a 10× magnification objective lens (Nikon LU Plan Fluor 10×, NA 0.30) was used in bright field mode, illuminated by a 50 W halogen lamp (Nikon LV-HL50PC). All microscope apertures were fully opened to maximize sample illumination. White balance calibration was performed using a white paper, with RGB component coefficients set to Rw = 1.43, Gw = 1.00, and Bw = 2.13 for color measurement experiments. Exposure time was set to 4 ms with a gain factor of 1.00, initially determined in auto-exposure mode and later applied manually. RGB images were captured at an 8-bit color depth, with a 640 × 480-pixel resolution in BMP format, achieved by averaging a 4 × 4 pixel area from the CCD into one pixel. The actual size of each imaged area was 0.87 × 0.65 mm²

## 4. Digital image processing

The digital image processing and area fraction and threshold characterization were performed by using a symbolic and numeric computing environment and software (Maple 18, Maplesoft).

### 4.1. Area fraction evaluation

The digital image processing procedure used for area fraction calculations of laser-damaged Cu is presented in Figure 3.

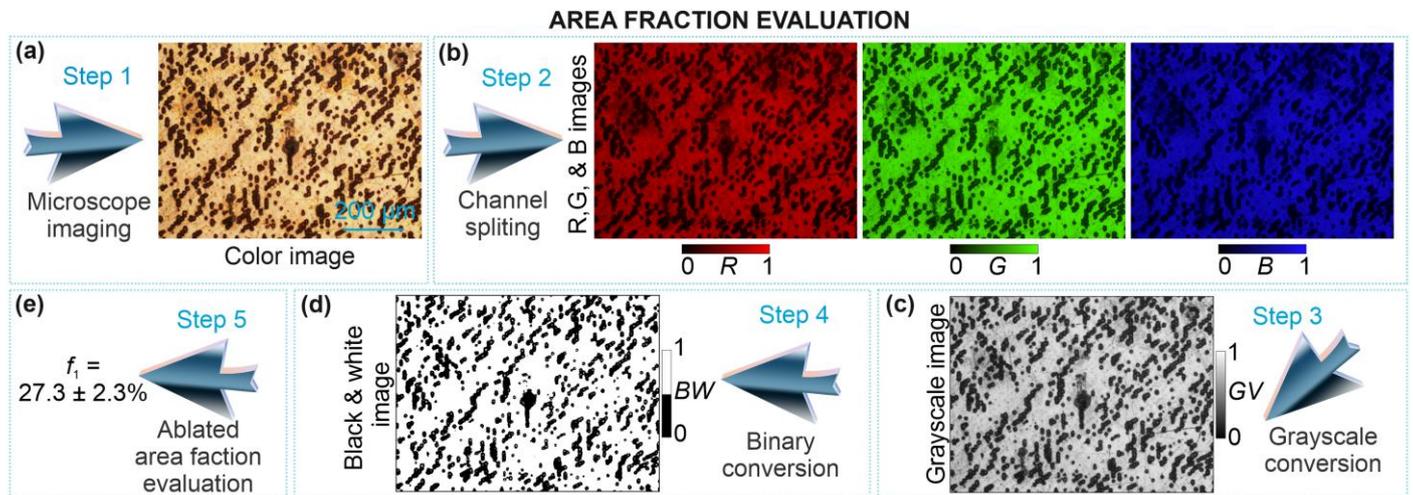

**Figure 3.** Digital image processing procedure designed for area fraction calculations: (a) step 1 – microscope imaging of laser-treated copper surface; (b) step 2 - color image split to red (R), green (G), and blue (B) channels; (c) step 3 - color-to-grayscale conversion by Eq. (6); (d) step 4 - grayscale image conversion to a black-and-white binary image using Eq. (7); (e) step 5 - calculation of the area fraction of $AF$ = 27.3 ± 2.3 % by Eq. (8).

In step 1 (Figure 3(a)) microscope color image of laser treated copper surface was taken. In step 2 (Figure 3(b)), the color image was split into red (R), green (G), and blue (B) channels. In step 3 (Figure

3(c)) the image was transformed to the grayscale mode by calculating the grayscale value $GV$ using the formula [99]:

$$GV = \max(R, G, B), \tag{6}$$

where $R$, $G$, and $B$ are the red, green, and blue components, respectively. The five color-to-grayscale conversion methods have been tested in our work: intensity, luminance, luma, luster, and value [7]. The value method has been chosen because of the highest achieved contrast between grayscale images of the laser-treated and untreated Cu. In step 4 (Figure 3(d)), the grayscale pictures were converted to black-and-white binary mode by using a certain threshold value $GV_{th}$ using formula [100]:

$$BW = \begin{cases} 1, & \text{if } GV \geq GV_{th}; \\ 0, & \text{if } GV < GV_{th}. \end{cases} \tag{7}$$

In step 5 (Figure 3(e)), the area fraction $f_1$ of the laser-treated copper Cu in percent was calculated by averaging the equation of binary image intensity [100]:

$$f_1 = \frac{1}{n}\sum_{i=1}^{n} BW_i \times 100\%, \tag{8}$$

where $n$ is the total number of picture pixels, $i$ is the pixel index, $BW_i$ is the binary black-and-white intensity of each pixel in the picture. The computational error in the area fraction $f_1$ evaluation was taken as the standard deviation of measurements from five sections of the microscope images.

### 4.2. Threshold evaluation

The digital image processing procedure used for the area threshold determination procedure is depicted in Figure 4.

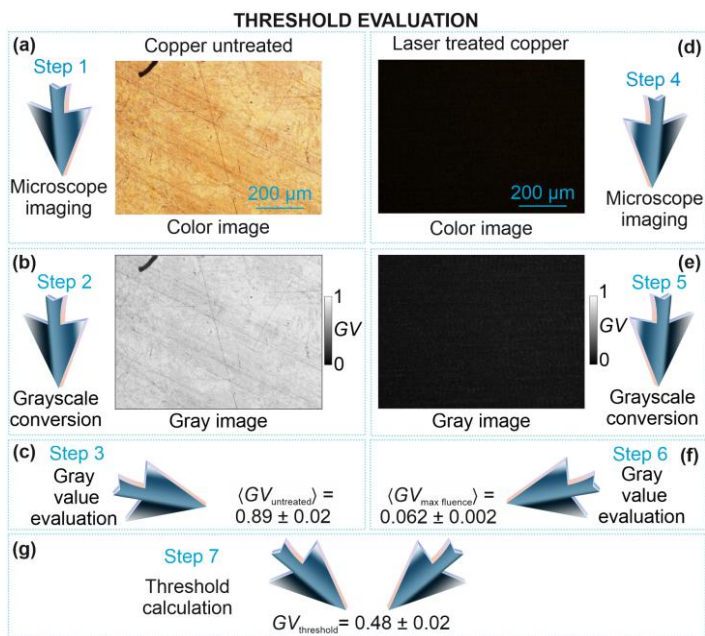

**Figure 4.** Threshold evaluation procedure: (a) step 1 – microscope imaging of laser-treated copper surface; (b) step 2 - color-to-grayscale conversion by Eq. (6); (c) step 3 - Average grayscale value calculation of untreated copper image of $\langle GV_{untreated}\rangle$ = 0.89 ± 0.02 by Eq. (9); (d) step 4 - microscope

imaging of laser treated copper surface at a max output power of 6.0 W with a corresponding max fluence of 9.6 J/cm²; (e) step 5 color-to-grayscale conversion by Eq. (6); (f) step 6 average grayscale value of $\langle GV_{\text{max fluence}}\rangle$ = 0.062 ± 0.002 evaluation by Eq. (9); (g) step 7 - threshold of $GV_{\text{th}}$ = 0.48 ± 0.02 calculation by Eq. (10).

In step 1 (Figure 4(a)) microscope image of untreated copper is taken. In step 2 (Figure 4(b)) the image is converted to the greyscale mode by using Eq. (6). In step 3 (Figure 4(c)) the averaged gray value of the image is evaluated by formula [100]:

$$\langle GV \rangle = \frac{1}{n}\sum_{i=1}^{n} GV_i, \tag{9}$$

where the average is denoted by angle brackets, $n$ is the total number of picture pixels, $i$ is the pixel index, $GV_i$ is the grayscale value of each pixel in the picture. The standard deviation as a computational error in the grayscale value $\langle GV \rangle$ evaluation was taken from five sections of the microscope images. The pictures were divided into 5 sections with a size of 128 × 480 pixels. The area fraction color was calculated for each section. The $\langle GV_{\text{untreated}}\rangle$ = 0.89 ± 0.02 of untreated copper (laser output power 0.0 W corresponding fluence 0.0 J/cm²) is calculated. In step 4 (Figure 4(d)) the microscope image of laser-treated copper at the maximum available laser output power of 6.0 W is taken. In step 5 (Figure 4(e)) the image is converted to the greyscale mode. In step 6 (Figure 4(f)) the averaged gray value of the image is evaluated by using Eq. (9). The $\langle GV_{\text{max fluence}}\rangle$ = 0.062 ± 0.02 of laser-treated copper (maximum laser output power 6.0 W and corresponding max fluence 9.6 J/cm²) is calculated. The threshold was selected as a mean of average values of untreated Cu (0.0 W) and laser-treated coper at maximum available laser power (6.0 W) [100]:

$$GV_{\text{threshold}} = \frac{GV_{\text{untreated}} + GV_{\text{max fluence}}}{2}. \tag{10}$$

In step 7 (Figure 4(g)) the threshold value calculated by Eq. (10) was $GV_{\text{threshold}}$ = 0.48 ± 0.02.

### 4.3. Luminance evaluation

The luminance $GL$ of grayscale was computed from RGB optical microscope images by using a formula based on the NTSC standard [99,101]:

$$GL = 0.3R + 0.59R + 0.11B. \tag{11}$$

where $R$, $G$, and $B$ are red, green, and blue components of the sample images after laser treatment.

### 4.4. Color distance evaluation.

The color distance $CD$ was calculated between digital optical microscope images of the copper surface before and after laser treatment was calculated by using equation [102]:

$$CD^2 = \left(R_{\text{untreated}} - R_{\text{fluence}}\right)^2 + \left(G_{\text{untreated}} - G_{\text{fluence}}\right)^2 + \left(B_{\text{untreated}} - B_{\text{fluence}}\right)^2. \tag{12}$$

where $R_{untreated}$, $G_{untreated}$, $B_{untreated}$, $R_{fluence}$, $G_{fluence}$, and $B_{fluence}$ are red, green, and blue components of the sample images before (untreated) and after laser treatment (at a certain value of laser fluence). The averaging procedure is performed by using equation (9). The average color difference ⟨CD⟩ and the standard deviation of it were calculated from the data achieved from five sections of images.

## 5. Results and discussion
### 5.1. Contact angle, color, and area fraction evaluation

The experimental results of the wettability and color change of copper depending on the laser fluence are depicted in Figure 5 (all experimental data is provided in supplementary material Figure S1).

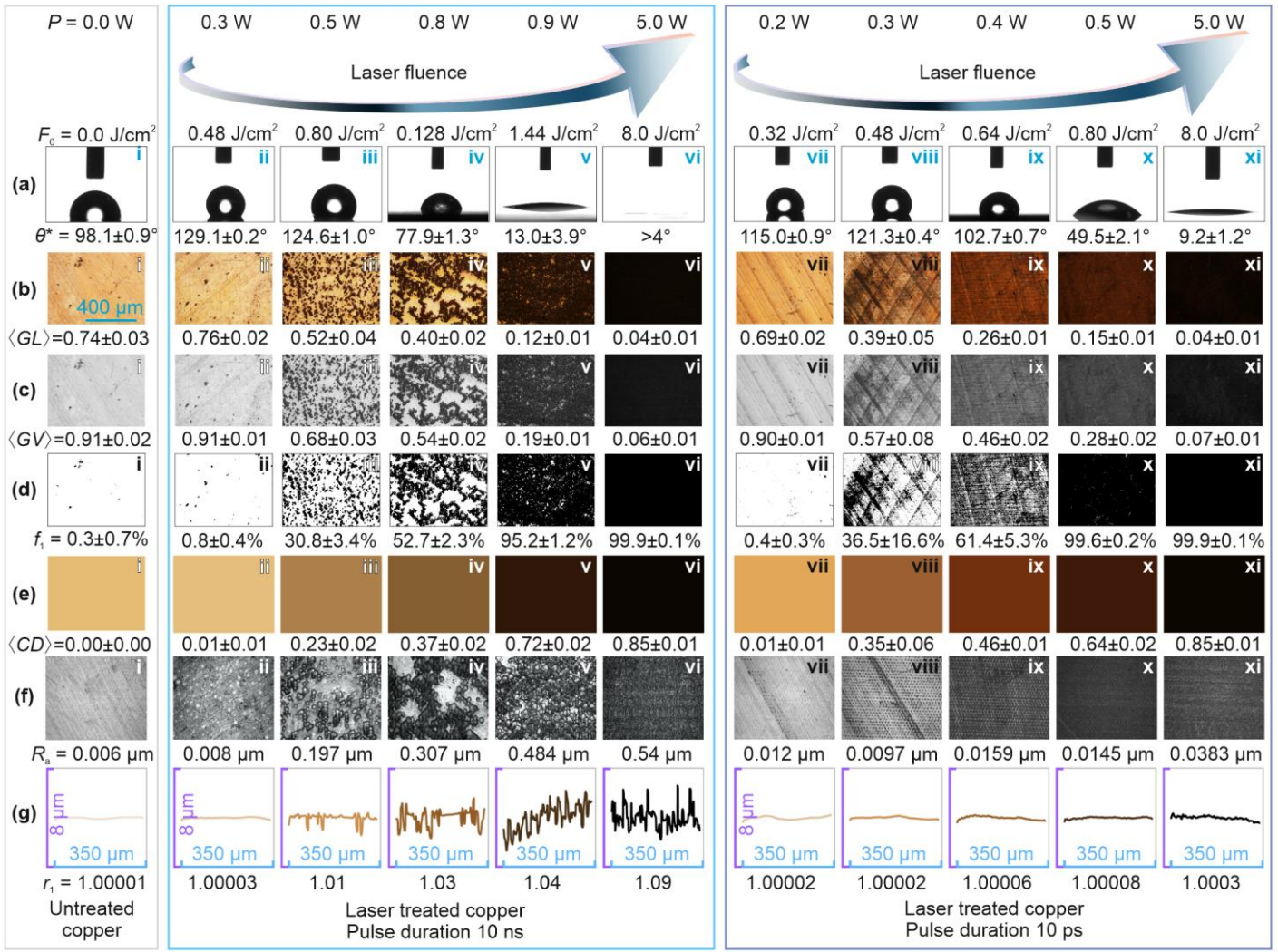

**Figure 5.** (a) Water droplet images on copper surface treated by laser at different laser fluences. (b) RGB color digital optical microscope images of the Cu surface. (c) Color optical microscope images converted to grayscale mode. (d) Grayscale images were converted to black-and-white binary images with the threshold *value* of $GV_{th}$ = 0.48 ± 0.02. (e) The reconstructed average color of color images of copper. (f) 3D optical profiler images of the copper surface. (g) Line profiles of laser-treated copper surface. The laser fluence used for treatment: (i) $F_0$ = 0.0 J/cm² (untreated); (ii) 0.48 J/cm²; (iii) 0.80 J/cm²; (iv) 0.128 J/cm²; (v) 1.44 J/cm²; (vi) 8.0 J/cm²; (vii) 0.32 J/cm² (viii) 0.48 J/cm² (ix)

0.64 J/cm² (x) 0.80 J/cm², and (xi) 8.0 J/cm². Pulse durations used (ii-vi) $\tau_p$ = 10 ns and (vii-xi) $\tau_p$ = 10 ps. The scale bar provided in (b)(i) is valid for all images in (b-d) rows.

Water droplet images on copper surfaces treated by laser at different laser fluences are provided in Figure 5(a). The measured contact angle $\theta^*$ values are given below each image. The contact angle decreased with increasing laser fluence for both nanosecond and picosecond laser irradiation regimes. The microscope RGB color images are provided in Figure 5(b). The average grayscale *luminance* $\langle GL \rangle$ calculated by Eq. (11) is given below each image. The size of each RGB microscope image is 0.87 × 0.65 mm². The RGB color image conversion to the grayscale mode is provided in Figure 5(c). The average grayscale *value* $\langle GV \rangle$ calculated by Eq. (9) is given below each image. The grayscale image conversion and black-and-white images are provided in Figure 5(d). The area fraction is calculated from black-and-white images by using Eq. (8) given below each image. The area fraction of the laser-ablated surface increases with increasing peak laser fluence for both nanosecond and picosecond laser irradiation regimes. The reconstructed color of microscope RGB images of the surface of copper is provided in Figure 5(e). The average color distance $\langle CD \rangle$ calculated by Eq. (12) is provided below each image. The grayscale luminance $\langle GL \rangle$, grayscale value $\langle GV \rangle$ and color distance $\langle GV \rangle$ of each image increased linearly with increasing peak laser fluence for both nanosecond and picosecond laser irradiation regimes. The 3D optical profiler images are provided in Figure 5(f). The surface roughness values $R_a$ is provided below each image. The surface roughness $R_a$ for both nanosecond and picosecond pulse duration increases with increasing laser fluence. However, the $R_a$ values for picosecond pulses are more than 10 times smaller than for nanosecond pulses. The line profiles of the laser-treated copper surface are provided in Figure 5(g) (all experimental data of line profiles is provided in the supplementary material Figure S2). The normalized profile arc length $r_1$ values calculated by Eq. (5) are provided below the profiles. The normalized profile arc length values increase for both nanosecond and picosecond pulse durations. However, the $r_1$ values for picosecond pulses are more than 100 times smaller than for the nanosecond pulses. It can be considered that laser-induced surface roughness and increased profile length for nanosecond pulses reduced the contact angles to values close to zero degrees [103]. However, for picosecond pulses the roughness is more than 10 times and the profile arc length is more than 100 times lower than for the nanosecond pulses. The contact angle also decreases to zero degrees, the Wenzel model can not be directly applied for the interpretation of the experimental results.

## 5.2. Contact angle and color vs area fraction

The plots depicting how different surface characteristics (contact angle, color distance, gray value, and gray luminance) vary with the ablated area fraction for two different laser pulse durations: $\tau_p$ = 10 ns and $\tau_p$ = 10 ps is depicted in Figure 6 (all experimental data is provided in the supplementary material Figure S3, Figure S4, and Figure S5).

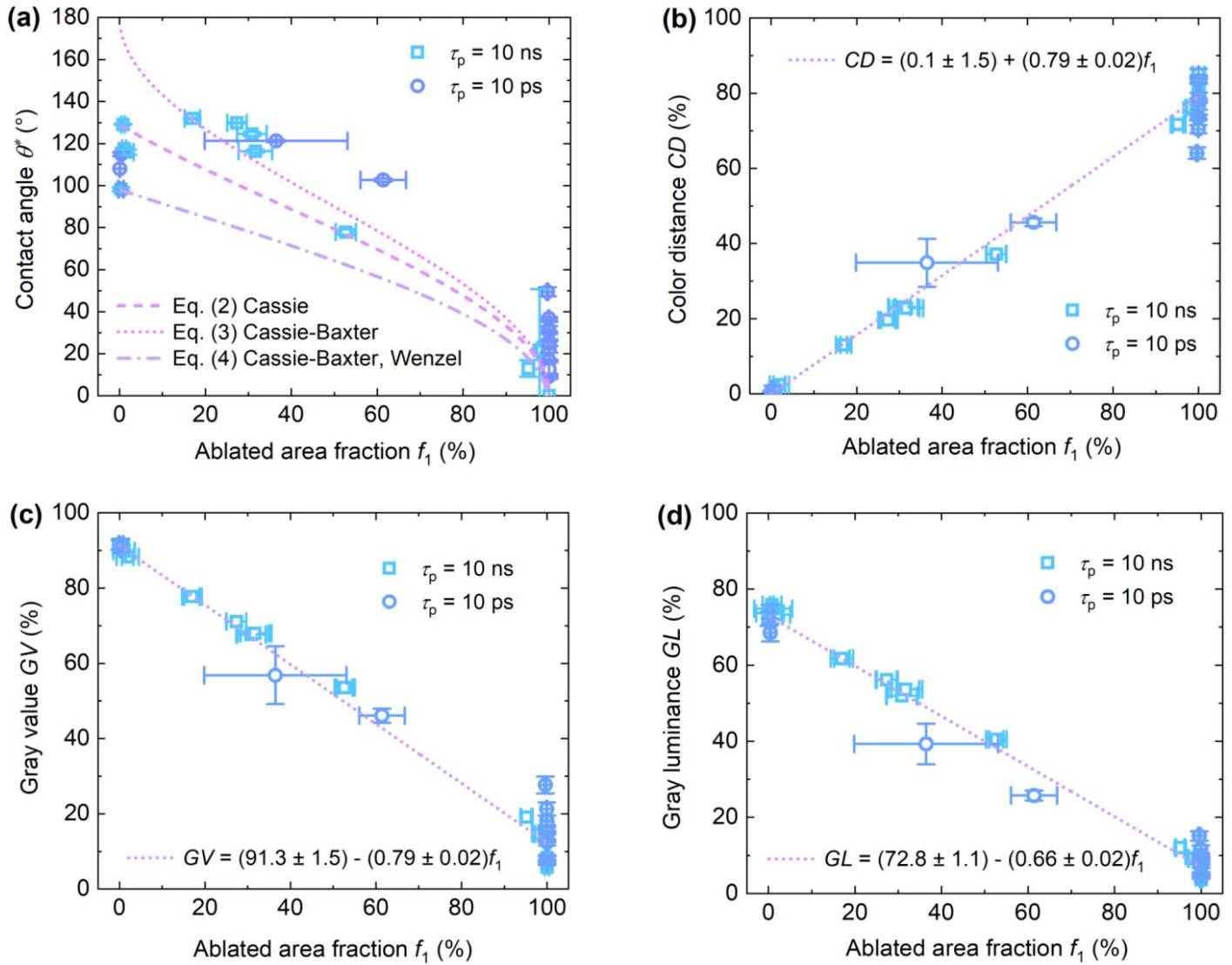

**Figure 6.** Water static contact angle $\theta^*$ (a), color distance $CD$ (b), gray value $GV$ (c), and gray luminance $GL$ (d) depend on the area fraction $f_1$ of laser-ablated copper. The open squares and open circles correspond to laser processing using irradiation at different pulse durations of $\tau_p$ = 10 ns and $\tau_p$ = 10 ps, respectively. The dot, dash, and dot-dash lines in (a) are fits of the experimental data point by Eq. (2), Eq. (3), and Eq. (4), respectively. The horizontal error bars in (a-d) indicate the standard deviation in the area fraction measurements taken from five sections of the microscope images. Vertical error bars in (a) correspond to the difference in contact angle measured on the left and right sides of the droplet. Vertical error bars in (b-d) indicate the standard deviation in the color distance, gray value, and gray luminance evaluation from five sections of the microscope images. Common processing conditions: irradiation wavelength $\lambda$ = 1064 nm, spot size on the sample $w_0$ = 20 µm; pulse repetition rate $f_{rep}$ = 100 kHz, beam scanning speed on the sample $v_{scan}$ = 1.0 m/s, the lateral distance between laser pulses $\Delta x$ = 10 µm, the lateral distance between bidirectional scanned lines in snake-like beam

trajectory Δy = 5 μm. The water contact angle on laser-textured surfaces was measured 15 days after laser processing.

The contact angle decreases as the ablated area fraction increases, meaning the surface becomes more hydrophilic (Figure 6(a)). The theoretical models provide predictions for how the contact angle should change based on different surface states (e.g., Cassie or Cassie-Baxter, representing different wettability regimes). Fitting parameters for Eq. (2), Eq. (3), and Eq. (4) for curves given in Figure 6(a) are depicted Table 1.

**Table 1** Fitting parameters used Figure 6(a).

| Contact angle $\theta_1$ | Contact angle $\theta_2$ | Profile arc length $r_1$ | Equation | Model | Reference |
| --- | --- | --- | --- | --- | --- |
| 0° | 129° | - | (2) | Cassie | 91,92 |
| 0° | - | - | (3) | Cassie-Baxter | 93,94 |
| 4° | 98° | 1.01 | (4) | Cassie-Baxter, Wenzel | 94,95 |

For Cassie model Eq. (2) the contact angles of $\theta_1 = 0°$ which correspond well to the experimental value minimal contacted angle of <4° of laser structured copper at high fluence (8 J/cm²) (Figure 5(a)(vi)), and $\theta_2 = 129°$ correspond well to the experimental value maximal measures angle of laser structured copper at low fluence (0.48 J/cm²) (Figure 5(a)(ii)).

For Cassie-Baxter model Eq. (3) the contact angles of $\theta_1 = 0.0°$ which correspond well to the experimental value minimal contacted angle of <4° of laser structured copper at high fluence (8 J/cm²)

For the combined Cassie-Baxter and Wenzel model Eq. (4) the contact angles of $\theta_1 = 4°$ which correspond well to the experimental value minimal contacted angle of <4° of laser structured copper at high fluence (8 J/cm²) (Figure 5(a)(vi)), and $\theta_2 = 98°$ correspond well to the experimental value maximal measures angle of laser unstructured copper (0.0 J/cm²) (Figure 5(a)(i)).

There is good agreement between experimental data and theoretical fits, especially for the $\tau_p = 10$ ps pulses (Figure 6(a)). The color change increases linearly with the ablated area fraction. A linear fit is provided: $CD = (0.1\pm1.5) + (0.79\pm0.02)f_1$ (Figure 6(a)). The $\tau_p = 10$ ps pulses appear to cause a more rapid color change compared to $\tau_p = 10$ ns pulses at low ablation fractions. (Figure 6(b)). The gray value decreases as the ablated area fraction increases, suggesting that the surface becomes darker as more material is ablated. The linear fit equation is: $GV = (91.3\pm1.5) - (0.79\pm0.02)f_1$ Similar trends are observed for both $\tau_p = 10$ ns and $\tau_p = 10$ ps, though there may be subtle differences in behavior at low ablation fractions (Figure 6(c)). Gray luminance decreases as the ablation fraction increases, consistent with the gray value trends. The linear fit is $GL = (72.8\pm1.1) - (0.66\pm0.02)f_1$. Both $\tau_p = 10$ ns and

$\tau_p$ = 10 ps pulses show similar trends, with a gradual reduction in luminance as ablation increases (Figure 6 (e)).

## 6. Conclusions

To conclude, here we reported on a novel single-step and chemical-free fabrication method for the creation of super-wetting and highly hydrophobic copper surfaces using nanosecond and picosecond lasers. The wettability and color of copper were controlled by controlling the area fraction of the laser-ablated surface. Our adopted Cassie, Cassie-Baxter, and Wenzel models indicate that as the ablated area fraction increases, the contact angle decreases, indicating that the surface becomes more hydrophilic. Both color distance, gray value, and gray luminance increase or decrease linearly with the ablated area fraction of the copper surface, indicating significant visual and optical changes due to surface modification. The pulse duration (ps or ns) plays a role in the degree of these changes, with shorter pulse durations (10 ps) typically having a stronger effect. Thanks to the laser-texturing process, super-wetting surfaces can be efficiently scaled up to cover large areas, even on complex shapes. This scalability makes them ideal for enhancing fog harvesting in atmospheric water generators and improving the performance of heat exchanger technologies used in water heat sinks, fans, and cooling units.

**Disclosures.** The authors declare no conflicts of interest.

**Data availability.** Data underlying the results presented in this paper are not publicly available at this time but may be obtained from the authors upon reasonable request.

**Acknowledgments.** M.Ga., A.Ž., and M.Ge. received funding from the Research Council of Lithuania (LMTLT), agreement no. S-MIP-22-89. This work was also supported by the EU-Horizon 2020 Nanoscience Foundries and Fine Analysis (NEP) Project (Grant agreement ID 101007417) having benefited from the access provided by the Foundation for Research and Technology Hellas (FORTH) (Access project ID 177).

**Funding.** Research Council of Lithuania (LMTLT) (S-MIP-22-89). This project has received funding from the European Union's Horizon 2020 research and innovation programme, under grant agreement No 101007417 NFFA-Europe Pilot.

**Contributions**
M.Ga., A.Ž., and M.Ge. conceived the original idea for the project. E.St. supervised for collaboration work and internship of M.Ge. and A.Ž. at IESL-FORTH. E.Sk. hosted the stay of M.Ge. and A.Ž. at IESL-FORTH.

A.L. helped with software development and A.P. helped with technical support at IESL-FORTH. S.M. and A.M. performed the water contact angle measurements at IESL-FORTH. M.Ge. performed the structuring using nanosecond and picosecond lasers at LTS-FTMC. M.Ga. conducted the surface topography measurements and surface roughness measurements using a 3D optical profiler at LTS-FTMC. M.Ge. supervised all the findings of this work, adopted the Casei-Baxter and Wenzel models, performed the digital image analysis, and numerical calculations, analyzed the data, and wrote the manuscript. All authors analyzed the findings, discussed the results, and commented on the manuscript.

**Supplementary material**

**Wettability and Color Change of Copper by Controlling Area Fraction of Laser Ablated Surface**

Mantas Gaidys[1], Stella Maragkaki[2], Alexandros Mimidis[2], Antonis Papadopoulos[2], Andreas Lemonis[2], Evangelos Skoulas[2], Andrius Žemaitis[1], Emmanuel Stratakis[2], Mindaugas Gedvilas[1]*

[1]Department of Laser Technologies (LTS), Center for Physical Sciences and Technology (FTMC), Savanoriu Ave. 231, 02300 Vilnius, Lithuania
[2]Institute of Electronic Structure and Laser (IESL), Foundation for Research and Technology (FORTH), N. Plastira 100, Vassilika Vouton, 70013 Heraklion, Crete, Greece

E-mail: mindaugas.gedvilas@ftmc.lt

The full set of experimental data of the wettability and color change of copper surface depending on the laser power is depicted in Figure S1.

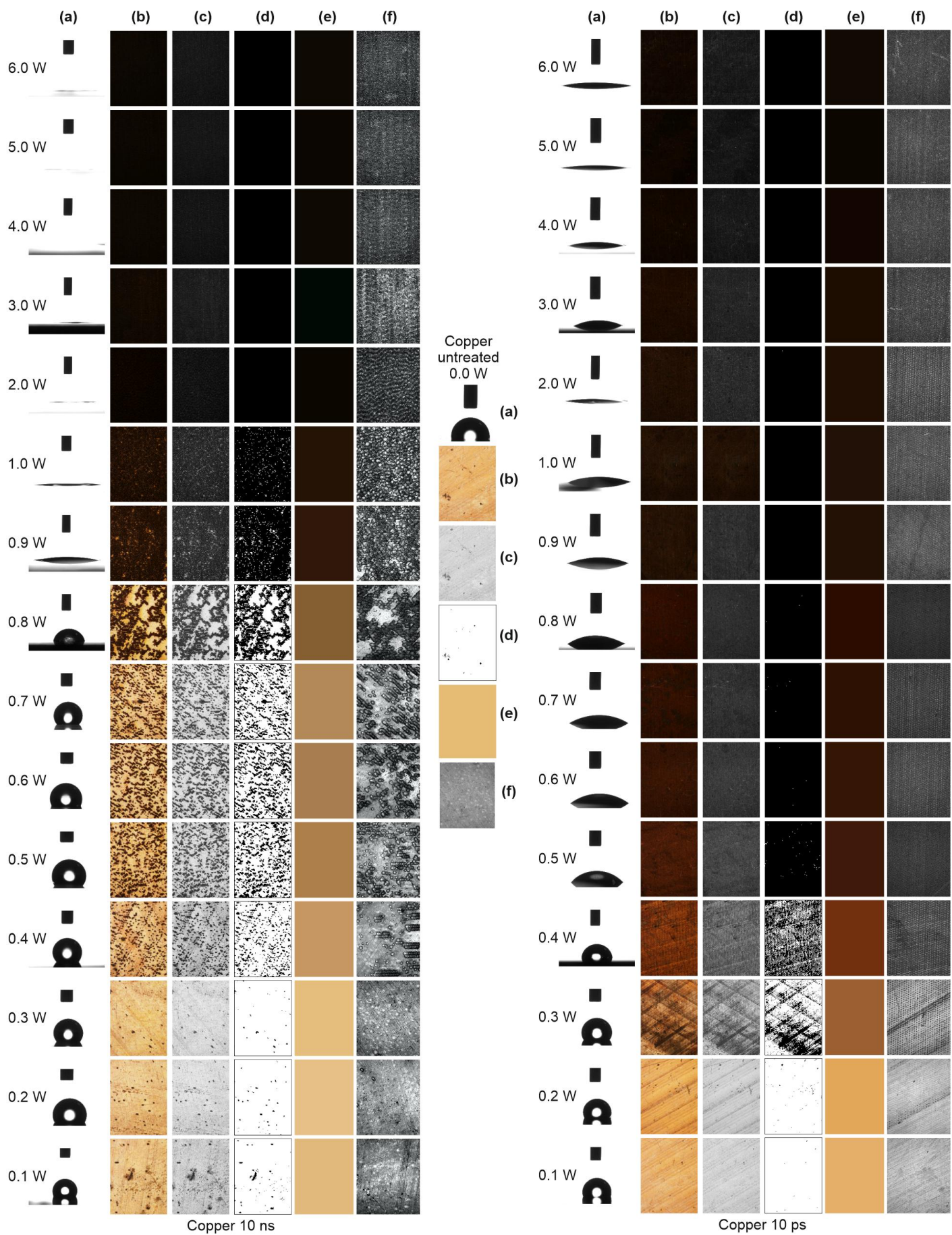

**Figure S1.** (a) Water droplet images on copper surface treated by laser at different laser powers from 6.0 W (top) to 0.1 W (left), untreated copper 0.0 W (middle). (b) RGB color digital optical microscope images of the Cu surface. The size of each microscope image is 0.65 × 0.87 mm². (c) The color optical microscope images are converted to grayscale mode. (d) The grayscale images were converted to black-

and-white binary images. (e) The reconstructed average color of color images of laser-treated copper. (f) 3D optical profiler images of the copper surface. Pulse durations 10 ns (left) and 10 ps (right). Untreated copper (0.0 W) (middle).

The height line profiles of the copper surface structured by laser at different laser powers are given in Figure S2.

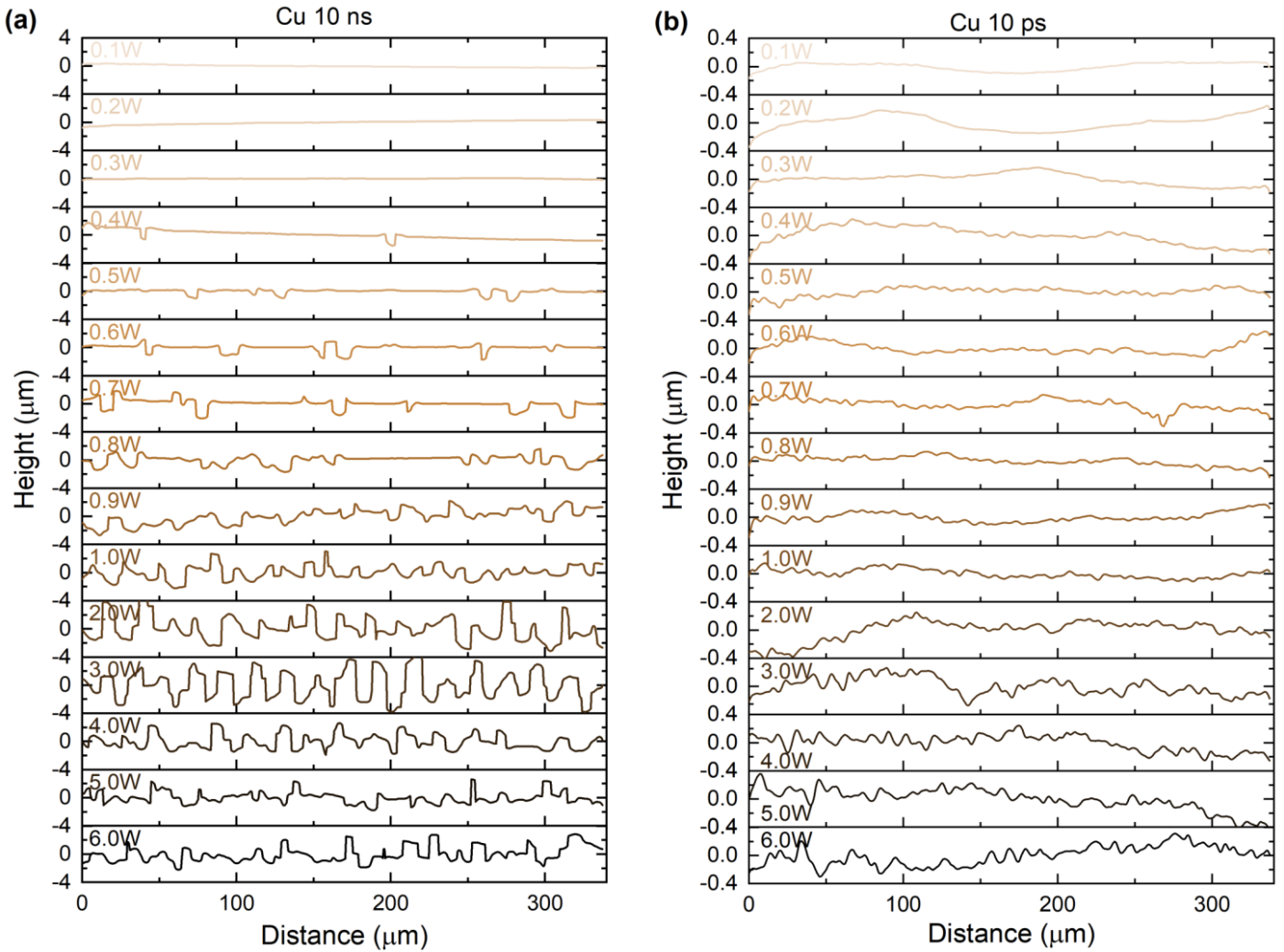

**Figure S2.** Height of line profiles dependence on transverse distance for copper structured by nanosecond (pulse duration $\tau_p$ = 10 ns) (a) and picosecond (pulse duration $\tau_p$ = 10 ps) (b) pulses at different laser irradiation powers from 0.1 W (top) to 6.0 W (bottom). Processing conditions: irradiation wavelength $\lambda$ = 1064 nm, spot size on the sample $w_0$ = 20 µm; pulse repetition rate $f_{rep}$ = 100 kHz, beam scanning speed on the sample $v_{scan}$ = 1.0 m/s, the lateral distance between laser pulses $\Delta x$ = 10 µm, the lateral distance between bidirectional scanned lines in snake-like beam trajectory $\Delta y$ = 5 µm.

Figure S2 (a) shows the surface profiles of copper after being exposed to 10 ns pulse duration laser at increasing power (from 0.1 W to 6.0 W). At low power (0.1–0.6 W) there are minimal height variations, with very small surface undulations close to the baseline of 0 µm. At moderate power (0.7–1.0 W) there are more pronounced height variations, with noticeable surface features such as shallow peaks and valleys. Ablation effects begin to appear at higher powers. At high power (2.0–6.0 W) the surface roughness increases dramatically. The profiles show deep valleys and sharp peaks, suggesting significant material removal and possible melting or vaporization. The surface appears to be much rougher, with height fluctuations approaching ± 4 µm.

Figure S2(a) shows the surface profiles of copper after exposure to 10 ps pulse duration laser at increasing power (from 0.1 W to 6.0 W). At low powers (0.1–0.6 W) surface height remains close to 0 µm, with only very minor undulations, similar to the 10 ns pulse case. At moderate powers (0.7–1.0 W) the height variations are still relatively small, but they are smoother compared to the 10 ns case. The surface remains largely intact, with some minor ablation features appearing at higher powers. At high powers (2.0–6.0 W) the surface starts showing more pronounced features as the power increases, but the ablation appears smoother and more controlled compared to the 10 ns pulses. The height variations remain within the range of - 0.4 µm to + 0.4 µm, indicating much less material removal or surface disruption compared to the nanosecond pulses.

The 10 ns pulse duration (Figure S2(a)) leads to much more aggressive surface modifications, with deeper and more chaotic ablation features, especially at higher powers. The surface roughness increases significantly as power rises. The 10 ps pulse duration (Figure S2(b)) results in more controlled and subtle changes to the surface, with smaller ablation depths and smoother profiles even at higher powers. This indicates that shorter pulses (picosecond) tend to produce more precise ablation, while longer pulses (nanosecond) result in more widespread surface damage.

Water contact angle and ablated area fraction dependence on the laser irradiation power for copper structured by nanosecond and picosecond pulses is depicted in Figure S3.

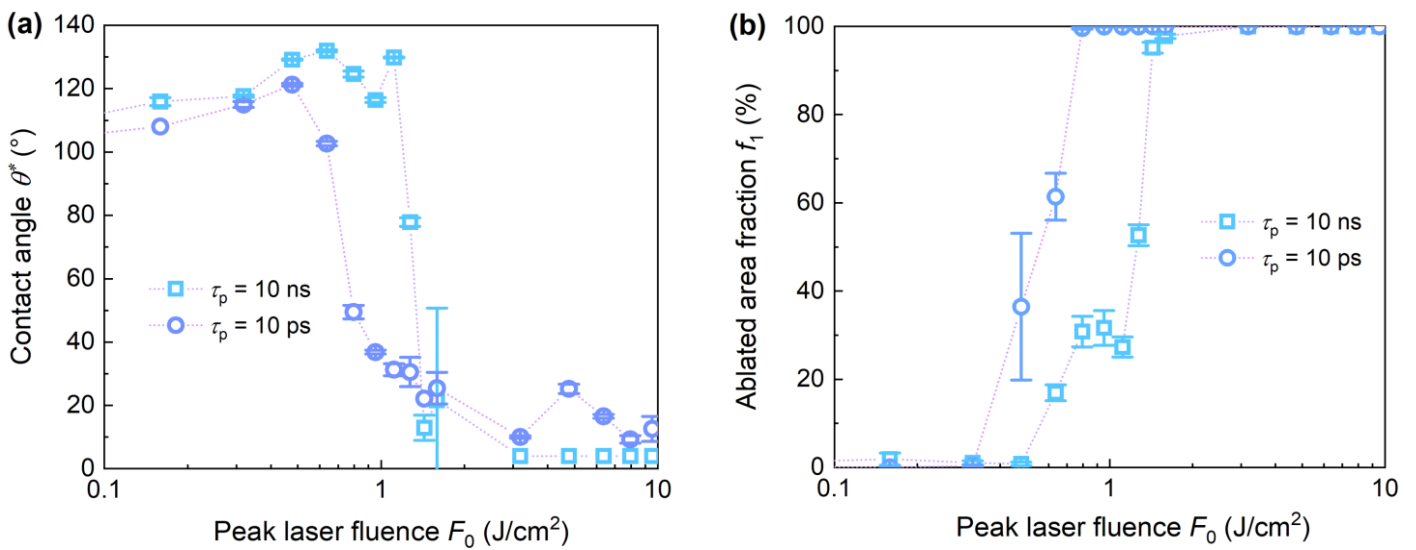

**Figure S3.** (a) Contact angle $\theta^*$ dependence on peak laser fluence $F_0$. (b) Ablated area fraction $f_1$ dependence on peak laser fluence $F_0$. Common processing conditions: irradiation wavelength $\lambda$ = 1064 nm, spot size on the sample $w_0$ = 20 µm; pulse repetition rate $f_{rep}$ = 100 kHz, beam scanning

speed on the sample $v_{scan}$ = 1.0 m/s, the lateral distance between laser pulses $\Delta x$ = 10 μm, the lateral distance between bidirectional scanned lines in snake-like beam trajectory $\Delta y$ = 5 μm. Laser pulse duration: $\tau_p$ = 10 ns - open squares; $\tau_p$ = 10 ps - open circles. The water contact angle on laser-textured surfaces was measured 15 days after laser processing.

At low fluences (~0.1–0.5 J/cm²), the contact angle remains high, indicating minimal changes in surface hydrophobicity. At fluences between 0.5 J/cm² and 1 J/cm², there is a sharp drop in contact angle, particularly for the $\tau_p$ = 10 ns. For both pulse durations, the contact angle stabilizes at lower values (~10°–20°) as fluence increases beyond 1 J/cm², indicating a transition to a more hydrophilic surface (Figure S3(a)). Contact angle (hydrophobicity) decreases as laser fluence increases, likely due to surface modification, which makes the surface more hydrophilic.

At fluences below 1 J/cm², the ablated area fraction is minimal for both pulse durations. There is a steep increase in ablated area fraction near 1 J/cm². For $\tau_p$ = 10 ps, the ablation onset occurs slightly earlier and results in a higher ablated fraction compared to $\tau_p$ = 10 ns (Figure S3(b)). Ablation begins around 1 J/cm², and the ps pulses (10 ps) lead to a higher ablation fraction than ns pulses (10 ns), suggesting that the shorter pulse duration is more effective at removing material.

Normalized profile arc length and surface roughness $R_a$ of copper structured by laser depending on peak laser fluence are depicted in Figure S4.

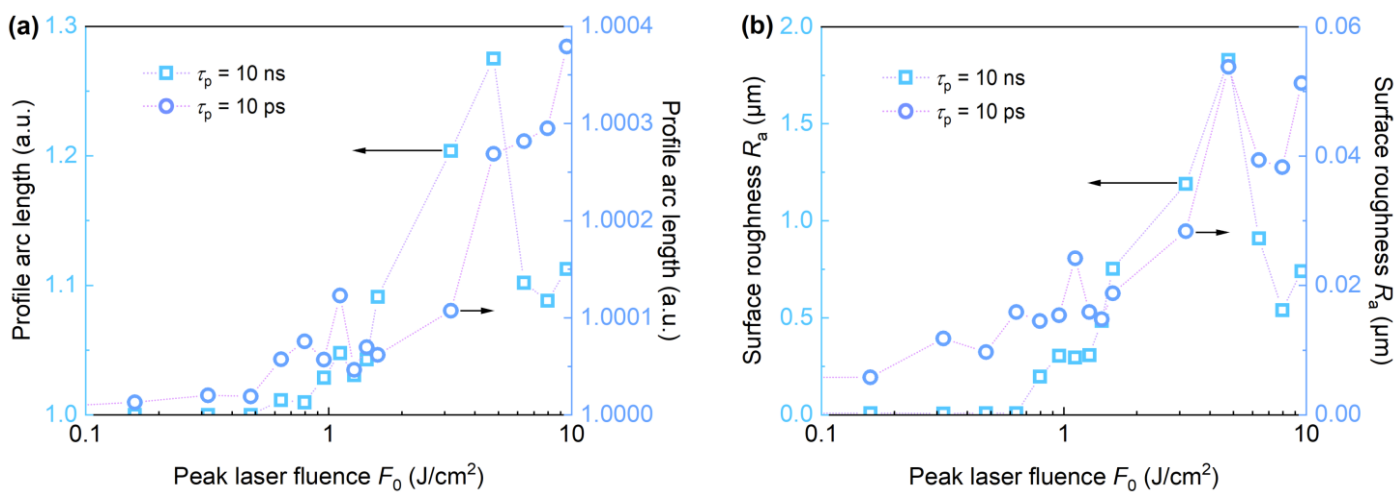

**Figure S4.** Normalized arc length of height profile line (a) and surface roughness $R_a$ (b) dependence on the peak laser fluence of copper surface structured by nanosecond ($\tau_p$ = 10 ns - open squares) and picosecond ($\tau_p$ = 10 ps - open circles) pulses.

Figure S4 illustrates the relationship between peak laser fluence profile arc length and surface roughness. The profile arc length increases as the peak laser fluence increases for both pulse durations, but the behavior is more pronounced at higher fluences (Figure S4(a)). The curve for $\tau_p$ = 10 ps shows a more gradual increase compared to the $\tau_p$ = 10 ns case, especially as the fluence exceeds 1 J/cm². At low fluence <1 J/cm² both curves are relatively flat, showing minimal change in arc length (Figure S4(a)).

Surface roughness increases as the laser fluence increases for both pulse durations, but the behavior is more significant for $\tau_p$ = 10 ns (Figure S4(b)). The roughness is relatively low (less than 0.5 μm) at low fluences (<1 J/cm²), but it spikes significantly for $\tau_p$ = 10 ns around 10 J/cm², reaching values as high as 2 μm. The $\tau_p$ = 10 ps, curve also shows an increase but at a much slower rate compared to the $\tau_p$ = 10 ns curve.

Normalized profile arc length and surface roughness $R_a$ of copper structured by laser depending on static water contact angle are depicted in Figure S5.

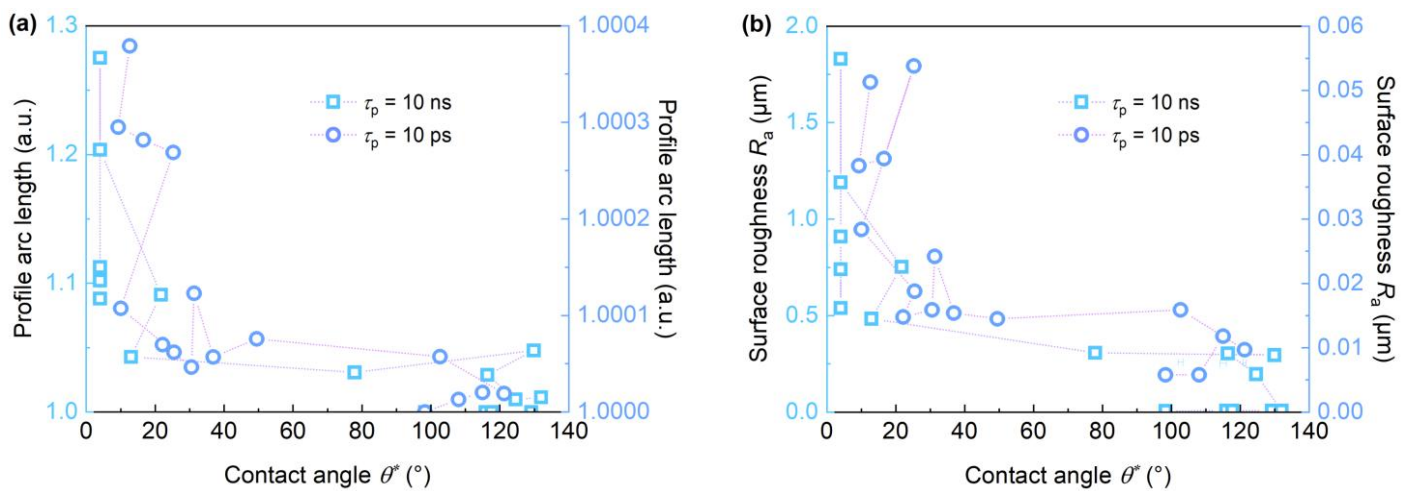

**Figure S5.** Normalized arc length of height profile line (a) and surface roughness $R_a$ (b) dependence on the static water contact angle of copper surface structured by nanosecond ($\tau_p$ = 10 ns - open squares) and picosecond ($\tau_p$ = 10 ps - open circles) pulses.

Figure S5 shows how profile arc length and surface roughness relate to the contact angle $\theta^*$, for two laser pulse durations. The graphs present how these surface characteristics change with varying contact angles, which reflects material wettability and texture, in response to laser processing.

The profile arc length decreases sharply as the contact angle increases from 0° to around 40° for both pulse durations Figure S5(a). For ns pulses, the profile arc length starts around 1.2 and decreases more drastically than for ps pulses. After the initial drop, both curves flatten, and the profile arc length stabilizes in the value range of 1.0 - 1.1 for contact angles greater than 60°. The curve for ps pulses is smoother compared to the ns pulses, showing a more gradual transition between contact angles.

For ns pulses, roughness starts near 2 μm at low contact angles (0°–10°) and decreases sharply to below 1 μm as the contact angle approaches 40° Figure S5(b). For ps pulses, the decrease is less dramatic but follows a similar trend, starting at around 1.5 μm and flattening after 40°. At higher contact angles (60° to 140°), surface roughness stabilizes at low values (below 0.5 μm), with slight fluctuations but no significant changes for either pulse duration.